\documentclass{aa}
\usepackage{graphicx}

\begin{document}

\title{A SCUBA survey of the NGC~2068/2071 protoclusters }

\author{F. Motte
        \inst{1}
        \thanks{\emph{Present address:} California Institute of 
        Technology, MS 320-47, Pasadena, CA 91125, USA}
   \and P. Andr\'e
        \inst{2}
   \and D. Ward-Thompson
        \inst{3}
   \and S. Bontemps
        \inst{4}}

\institute{Max-Planck-Institut f\"ur Radioastronomie, Auf dem H\"ugel 
           69, 53121 Bonn, Germany
      \and CEA/DSM/DAPNIA, Service d'Astrophysique, C.E.~Saclay, 91191 
           Gif-sur-Yvette Cedex, France
      \and Department of Physics \& Astronomy, University of Cardiff,
           P.O. Box 913, Cardiff, UK
      \and Observatoire de Bordeaux, BP 89, 33270 Floirac, France }

\offprints{F. Motte or P. Andr\'e, \email{motte@submm.caltech.edu, pandre@cea.fr}}
\date{Received 18 January 2001 / Accepted 12 April 2001}

\abstract{We report the results of a submillimeter dust continuum 
survey of the protoclusters NGC~2068 and NGC~2071 in Orion~B carried
out at $850\:\mu$m and $450\:\mu$m with SCUBA on JCMT.  The mapped
region is $\sim 32\arcmin \times 18\arcmin$ in size ($\sim
4$~pc~$\times$~2~pc) and consists of filamentary dense cores which
break up into small-scale ($\sim 5\,000$~AU) fragments, including 70
starless condensations and 5 circumstellar envelopes/disks.  The
starless condensations, seen on the same spatial scales as
protostellar envelopes, are likely to be gravitationally bound and
pre-stellar in nature.  Their mass spectrum, ranging from $\sim
0.3~M_\odot$ to $\sim 5~M_\odot$, is reminiscent of the stellar
initial mass function (IMF).  Their mass-size relation suggests that
they originate from gravitationally-driven fragmentation.  We thus
argue that pre-collapse cloud fragmentation plays a major role in
shaping the IMF.
}

\authorrunning{F. Motte et al.}
\maketitle

\section{Introduction}

The question of the origin of the stellar initial mass function (IMF),
crucial for both star formation and Galactic evolution, remains a
matter of debate (e.g. Larson 1999; Elmegreen 2001).  Numerous
molecular line studies of cloud structure have established that the
mass spectrum of observed clumps is significantly shallower than the
IMF (see, e.g., Williams et al. 2000 and references therein).  The
reason for this difference is presumably that most of the clumps
detected in CO surveys are not gravitationally bound and reflect more
the characteristics of the low-density outer parts of clouds than the
initial conditions of protostellar collapse (cf. Kramer et al. 1998).

The recent advent of sensitive bolometer arrays on large
(sub)millimeter radiotelescopes has made possible extensive surveys of
nearby star-forming clouds for young protostars and their pre-stellar
precursors (see, e.g., review by Andr\'e et al. 2000).  Using the
MPIfR bolometer array on the IRAM 30~m telescope, Motte et al. (1998
-- hereafter MAN98) could identify a total of 58 small-scale
($1\,000-6\,000$~AU), gravitationally-bound starless condensations in
their 1.3~mm continuum mosaic of the $\rho$~Ophiuchi central cloud.
Remarkably, the mass distribution of these pre-stellar condensations,
which spanned the range $\sim 0.05$ to $\sim 3~M_\odot$, mimicked the
shape of the stellar IMF.  It followed approximately the Salpeter
power-law IMF, $\Delta N/\Delta M \propto M^{-2.35}$, above $\sim
0.5~M_\odot$, and flattened out to $\Delta N/\Delta M \propto
M^{-1.5}$ at low masses.  Interestingly, the position of the break
point at $\sim0.5~M_\odot$ was comparable to the typical Jeans mass in
the dense DCO$^+$ cores of $\rho$~Ophiuchi (cf. Loren et al. 1990).
The results of MAN98 have been essentially confirmed by a $850\:\mu$m
survey of the same region with SCUBA (Johnstone et al. 2000).\\
In a related 3~mm interferometric study with OVRO, Testi \& Sargent
(1998) identified 26 compact starless condensations between $\sim 0.5$
and $\sim 10~M_\odot$ in the Serpens main cloud core.  They measured a
$\Delta N/\Delta M \propto M^{-2.1} $ mass spectrum, again close to
the Salpeter IMF.

These recent results are very promising as they support the view that
the IMF is at least partly determined by fragmentation at the
pre-stellar stage of star formation.  However, they are still limited
by small-number statistics and need to be confirmed in other
star-forming clouds.\\
In this Letter, we report on a wide-field submillimeter continuum
survey of the NGC~2068/2071 region in the Orion~B cloud complex.
Orion~B, also called L~1630, is the nearest giant molecular cloud
forming high-mass stars (see review by Launhardt \& Lada 2001) in the
vicinity of the Sun ($d \sim 400$~pc).  This complex contains five
active regions where rich clusters of young stellar objects (YSOs) are
currently forming within CS dense cores (Lada et al. 1991a,b).  The
two northernmost protoclusters, associated with the reflection nebulae
NGC~2068 and NGC~2071, provide good targets to explore the mass
spectrum of pre-stellar condensations over a broad mass range.

\section{Observations and data analysis}

\begin{figure}	
\vspace{-0.6cm}
\hspace{-0.3cm}\includegraphics[width=9.3cm]{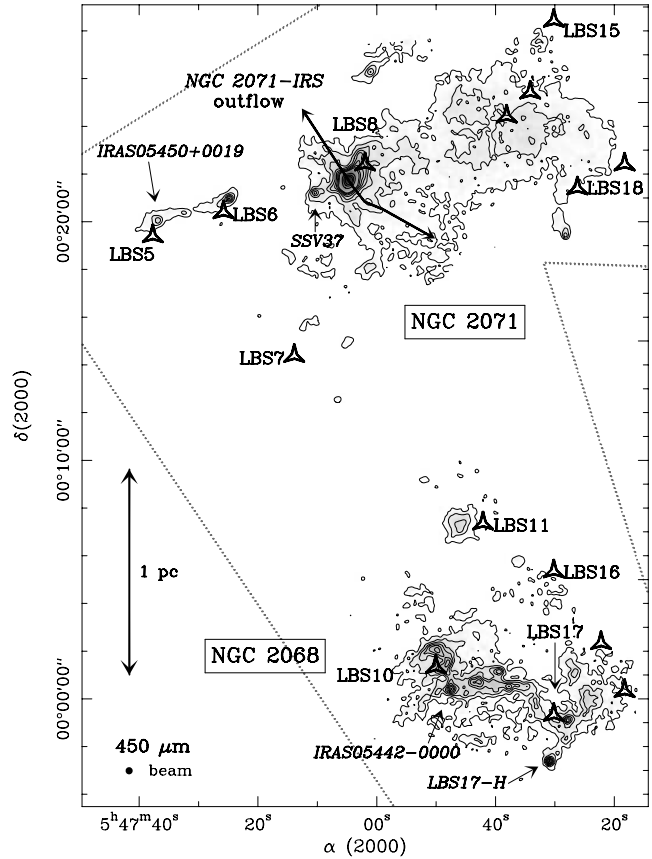}
\vspace{-1.5cm}	
\caption{Dust continuum mosaic of the NGC~2068/2071 region at 
$450\:\mu$m, smoothed to an effective angular resolution of
$18\arcsec$.  Contour levels go from 1.2 to 9.6~Jy/beam with steps of
1.2~Jy/beam and from 20 to 50~Jy/beam by 10~Jy/beam.  The mean rms
noise level is $\sim 0.4~\rm Jy/18\arcsec$-beam.  Several embedded
YSOs (italicized names) and CS cores (LBS numbers and triangles) are
indicated.}
\label{mosaic}
\end{figure}

We used the Submillimetre Common User Bolometer Array (SCUBA --
Holland et al. 1999) on the James Clerk Maxwell Telescope
(JCMT\footnote{JCMT is operated by the JAC, Hawaii, on behalf of the
UK PPARC, the Netherlands OSR, and the Canadian NRC.})  on 1998
December 15 and 16 to carry out a submillimeter continuum mapping of
NGC~2068 and NGC~2071. Five sub-fields were imaged simultaneously at
$850\:\mu$m and $450\:\mu$m in the standard scan-map mode to produce
two mosaics of a $\sim 32\arcmin \times 18\arcmin$ field (see
Fig.~\ref{mosaic}). Each sub-field was covered twelve times using
three different chop throws in both right ascension and declination.
Pointing and calibration checks were made on HL~Tau at regular
intervals.  The zenith atmospheric optical depth was measured to be
$\sim 0.25$ at $850\:\mu$m and $\sim 1.4$ at $450\:\mu$m.  The {\it
FWHM} beam size as measured on Uranus was $\sim 13\arcsec$ at
$850\:\mu$m and $8\arcsec$ at $450\:\mu$m.  The mosaics were reduced
with a SURF script from R. Tilanus using the ``Emerson~2'' restoration
algorithm (Emerson 1995).

\begin{figure}
\vspace{-0.7cm}
\hspace{-0.5cm}\includegraphics[height=9.2cm,angle=270]{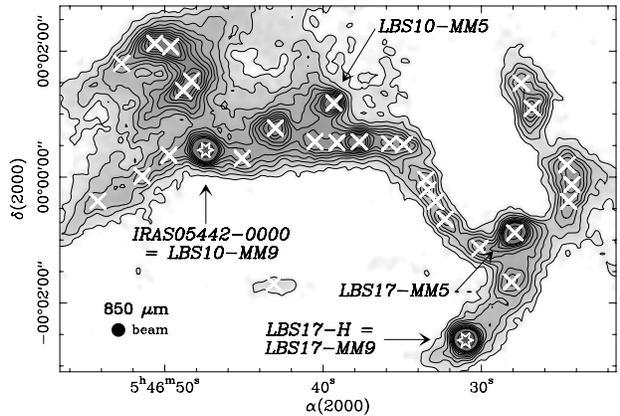}
\vspace{-0.3cm}
\caption{
Blow-up $850\:\mu$m continuum map of a sub-region around NGC~2068.
The effective angular resolution is $13\arcsec$.  Contour levels go
from 65 to 650~mJy/beam with steps of 65~mJy/beam; last contour is
1~Jy/beam.  The mean rms noise level is
$\sim$~22~mJy/$13\arcsec$-beam.  Starless condensations are denoted by
crosses and embedded YSOs by star markers.}
\label{condens}
\end{figure}

With a spatial extent of 3.7~pc~$\times$~2.1~pc, our mosaics cover
$75\%$ of the region mapped in CS(2--1) by Lada et al. (1991a --
hereafter LBS) around NGC~2068/2071.  In the submm continuum, the CS
dense cores of LBS have a filamentary appearance, with typical
dimensions $\sim $~0.08~pc~$\times$~0.5~pc (aspect ratio~$\sim 0.15$),
and are highly fragmented. Within these extended filaments, a total of
82 condensations with lengthscales characteristic of YSO circumstellar
structures (see Table~1, only available in electronic form at
http://cdsweb.u-strasbg.fr/A+A.htx, and Fig.~\ref{condens}) were
identified using a multiresolution wavelet analysis (cf. Starck et
al. 1998; MAN98; and Motte \& Andr\'e 2001).  The outer sizes of the
condensations were estimated from their radial intensity profiles at
$850\:\mu$m.  The ambient background observed on larger lengthscales
was then subtracted to allow a proper derivation of the integrated
flux densities of the condensations (see MAN98).  Unlike the
filaments, the (resolved) condensations are roughly circular (mean
aspect ratio~$\sim 0.6$) with an average deconvolved {\it FWHM}
diameter of $\sim 13\arcsec$ (i.e. $\sim 5\,000$~AU).

Our $850\:\mu$m mosaic (e.g. Fig.~\ref{condens}) mainly traces
optically thin thermal dust emission from cold, dense structures in
the cloud.  The map contours should thus primarily reflect the column
density distribution of dust and gas.  The condensation masses were
estimated from the integrated $850\:\mu$m fluxes (cf. MAN98), adopting
recommended values for the dust opacity per unit mass column density
(of dust {\it and} gas): $\kappa_{850} = 0.02~{\rm cm}^{2} \, {\rm
g}^{-1}$ for protostellar envelopes and $\kappa_{850} = 0.01~{\rm
cm}^{2} \, {\rm g}^{-1}$ for starless condensations (cf. Henning et
al. 1995).  The dust temperature $T_\mathrm{d} $ was taken to be
$20-40$~K and $15$~K, respectively, in agreement with published
temperatures (Harju et al. 1993; Gibb \& Little 2000).  Under these
assumptions, our 5$\sigma$ column density sensitivity is
$N_\mathrm{H2} \sim 1-2 \times 10^{22}$~cm$^{-2}$.  The condensations
have masses ranging from $\sim 0.3~M_\odot$ to $\sim 5~M_\odot$ ($\sim
9~M_\odot$ for NGC~2071-IRS), with a factor $\ga 2$ absolute
uncertainty.  The average density $\langle n_\mathrm{H2} \rangle$ of
the condensations is $3-40 \times 10^6~{\rm cm}^{-3}$ while that of
the filaments is $\sim 5 \times 10^5~{\rm cm}^{-3}$.

The $850\:\mu$m emission of seven condensations identified in the
vicinity of the luminous embedded infrared source NGC~2071-IRS
(e.g. Harvey et al. 1979) may not arise only from dust.  NGC~2071-IRS
(called LBS8-MM18 in Table~1) drives a prominent bipolar flow
(e.g. Chernin \& Masson 1992), which is responsible for a
northeast-southwest ridge of broad-band $850\:\mu$m emission.  The
spectral index observed between $450\:\mu$m and $850\:\mu$m in this
ridge is atypical: $\alpha \la 3$ (where $S_\nu \propto \nu^\alpha$)
instead of $\alpha \sim 4$ as measured for the extended emission
outside the ridge.  Based on the CO(3--2) and HCO$^+$(4--3) maps of
Chernin \& Masson (1992) and Girart et al. (1999), we estimate that
line emission may contribute up to $20-100\%$ of the SCUBA $850\:\mu$m
emission in the outflow region.  We are, however, confident that the
75 other condensations represent genuine dust continuum sources. Many
of them are detected at both $450\:\mu$m and $850\:\mu$m
(cf. Figs.~\ref{mosaic} and \ref{condens}).

\section{Characteristics of the condensations}

Among the 75 dust condensations of Sect.~2, four are associated with
both point-like mid-IR sources detected by ISOCAM (in the ISO core
programme -- e.g. Olofsson et al. 2000) and near-IR sources
(e.g. Strom et al. 1976).  These four objects most likely correspond
to circumstellar envelopes and/or disks around embedded YSOs.  In
addition, the condensation called LBS17-MM9 in Table~1 coincides with
the Class~0 outflow driving source LBS17-H of Gibb \& Little (2000).
The remaining 70 condensations appear to be truly starless.  Several
of them are seen in absorption by ISOCAM against the diffuse, ambient
mid-IR background (Bontemps et al. in prep.), which is reminiscent of
isolated pre-stellar cores (Bacmann et al. 2000).\\
Six submillimeter condensations (5 starless sources plus the Class~0
envelope LBS17-H) have been mapped in high-density molecular tracers
with good ($\la 20 \arcsec$) angular resolution (e.g. Gibb et
al. 1995).  Comparing the virial masses estimated by Gibb et al. from
HCO$^+$(3--2) observations with the masses derived here from the submm
continuum, we find $M_\mathrm{smm}/M_\mathrm{vir} \ga 0.2$. This
suggests that the condensations are close to virial equilibrium,
although more extensive spectroscopic observations in an optically
thin line tracer would be required to draw definitive conclusions.

\begin{figure}
\vspace{-0.5cm}
\hspace{-0.1cm}\includegraphics[height=9.3cm,angle=270]{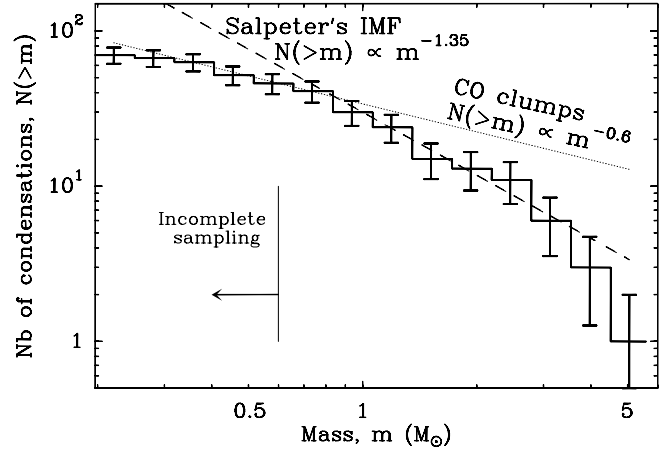}
\vspace{-0.1cm}
\caption{
Cumulative mass distribution of the 70 pre-stellar condensations of
NGC~2068/2071.  The dotted and dashed lines are power-laws
corresponding to the mass spectrum of CO clumps (Kramer et al. 1996)
and to the IMF of Salpeter (1955), respectively.  The error bars
correspond to $\sqrt N$ counting statistics. }
\label{specmass}
\end{figure}

Fig.~\ref{specmass} shows the cumulative mass spectrum ($N(>m)$
vs. $m$) of the 70 starless condensations identified in
NGC~2068/2071. The mass spectrum for the 30 condensations of the
NGC~2068 sub-region (cf. Fig. 2) is very similar in shape.  The
best-fit power-law is $N(>m) \propto m^{-1.1}$ above $0.8~M_\odot$,
which is close to the Salpeter IMF, $N(>m) \propto m^{-1.35}$.  A
flattening of the mass distribution to $N(>m) \propto m^{-0.5}$ is
apparent below $0.8~M_\odot$ in Fig.~\ref{specmass}.  The break point
is, however, close to our completeness limit at $m \sim 0.6~M_\odot$
($5\sigma$ detection level for the largest condensations with ${\rm
\it FWHM} \sim 30\arcsec$ -- see Fig.~4).  Altogether, the derived
mass spectrum is in good agreement with the IMF of field stars which,
in cumulative form, scales roughly as a $N(>m) \propto m^{-0.5}$ at
low masses ($0.1 \la m \la 0.5~M_\odot$) and steepens to $N(>m)
\propto m^{-1.5}$ for $0.5 \la m \la 10~M_\odot$ (e.g. Scalo 1998).
By contrast, this mass spectrum is much steeper than the $N(>m)
\propto m^{-0.6}$ power-law measured by LBS and Kramer et al. (1996)
in their CS and CO studies of Orion~B.  The latter is rejected by a
Kolmogorov-Smirnov test at the $96\%$ confidence level.

\begin{figure}
\vspace{-0.2cm}
\hspace{0.2cm}\includegraphics[height=8.1cm,angle=270]{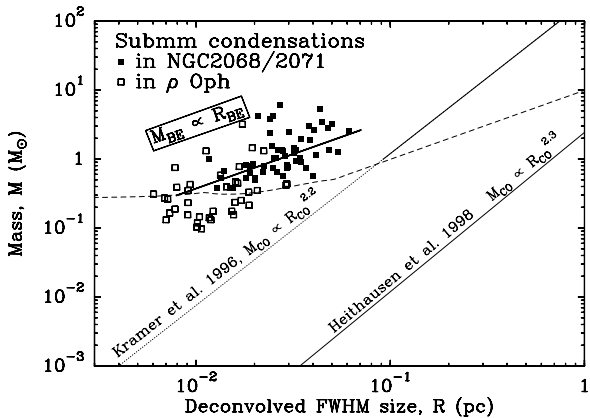}
\vspace{-0.2cm} 
\caption{
Mass-size relation of resolved (sub)mm condensations.  The solid line
is the relation expected for critical Bonnor-Ebert spheres with $T =
15 $~K and external pressures ranging from $P/k = 10^5$~K$\, \rm
{cm}^{-3}$ to $P/k = 8 \times 10^6$~K$\, \rm {cm}^{-3}$.  The dashed
curve shows our present $5\sigma $ detection threshold as a function
of size. The light lines show correlations observed for CO clumps.}
\label{MvsR}
\end{figure}

Fig.~\ref{MvsR} compares the mass-size relations derived for the
(sub)mm continuum condensations of the NGC~2068/2071 and $\rho$~Oph
protoclusters (from this paper and MAN98, respectively) with those
found for CO clumps in various clouds (e.g. Heithausen et al. 1998).
It can be seen that the (sub)mm continuum condensations are more than
one order of magnitude denser than typical CO clumps.  The mass-size
relation of the submillimeter condensations spans only one decade in
size and is much flatter than that of CO clumps: a formal fitting
analysis gives $M_\mathrm{smm} \propto (R_\mathrm{smm})^{1.1}$ as
opposed to $M_\mathrm{CO} \propto (R_\mathrm{CO})^{2.3}$.  Although
the observed correlations may be partly affected by size-dependent
detection thresholds (cf. Fig.~\ref{MvsR}), it is worth pointing out
that they are suggestive of a change from a turbulence-dominated to a
gravity-dominated regime.  Indeed, while the Larson law $M \propto
R^2$ is consistent with the fractal, turbulent nature of molecular
clouds (e.g. Elmegreen \& Falgarone 1996), a linear correlation ($M
\propto R$) is expected for a sample of self-gravitating isothermal
Bonnor-Ebert condensations assuming a uniform temperature and a range
of external pressures (cf. Bonnor 1956 and Fig.~\ref{MvsR}).  Most CO
clumps are transient structures associated with low density contrasts
and probably arise from hierarchical fragmentation driven by
turbulence (e.g. Elmegreen \& Falgarone 1996).  The starless
condensations identified here are much more centrally concentrated and
clearly require the additional effects of self-gravity.  Their
properties (e.g. Fig.~\ref{specmass} and Fig.~\ref{MvsR}) make them
excellent candidates for being the immediate progenitors of accreting
(Class~0/Class~I) protostars.
 
\section{Conclusions: Origin of the IMF in clusters}

The results of Sect.~3 suggest that the 70 starless condensations
identified in the NGC~2068/2071 protoclusters are about to form stars
on a one-to-one (or perhaps one-to-two) basis, with a high ($\ga
50\%$) efficiency roughly independent of mass.  They confirm the
findings of MAN98 and Testi \& Sargent (1998) in $\rho$~Ophiuchi and
Serpens.  There is now a growing body of evidence that the
fragmentation of dense ($\sim 10^5-10^6~{\rm cm}^{-3}$) cores into
compact, self-gravitating condensations determines the IMF of star
clusters in the low- to intermediate-mass range ($0.1-5\ M_\odot$).  A
plausible scenario, supported by some numerical simulations of cluster
formation (Klessen \& Burkert 2000; Padoan et al. 2001), could be the
following. First, cloud turbulence generates a field of density
fluctuations, a fraction of them corresponding to self-gravitating
fragments. Second, these fragments (or ``kernels'') decouple from
their turbulent environment (e.g. Myers 1998) and collapse to
protostars after little interaction with their surroundings.
 
The fact that none of the NGC~2068/2071 pre-stellar condensations is
found more massive than $5~M_\odot$ is statistically consistent with a
Salpeter-like mass distribution, given the relatively small number of
objects.  A more extensive submillimeter mapping of the Orion~B
complex should be done to improve the statistics and search for
starless condensations of higher mass.  However, massive stars may not
form from the collapse of single condensations but from the merging of
several pre-/proto-stellar condensations of low to intermediate mass.
In the collision scenario of Bonnell et al. (1998), the cluster
crossing time must be short enough to allow individual condensations
to collide and coalesce with one another.  Follow-up dynamical studies
of the NGC~2068/2071 condensations in dense molecular tracers could
help decide whether they have the potential to form massive stars with
$M_\star \ga 10~M_\odot$.

\begin{acknowledgements}
We would like to thank Jason Kirk for his participation during the
observing run.
\end{acknowledgements}


\begin{table*}[htbp]
\caption{Characteristics of the protocluster condensations identified in NGC2068/2071}
\begin{tabular}{|lcc|rcr|rl|}
\hline
 Condensation & \multicolumn {2} {c} {Coordinates}~$^{(1)}$ &
 $S_{850}^\mathrm{~int}$~$^{(2)}$ & {\it FWHM}~$^{(3)}$ & $M$~$^{(4)}$
 & $\alpha_{450}^{850}$ & Comments\\ Name & $\alpha_{2000}$ &
 $\delta_{2000}$ & [mJy] & [AU$~\times~$AU] & [$M_\odot$] & (5) & \\
\hline\hline
LBS18-MM1   & 05:46:28.2 &  00:19:29 &$1\,510$& $5\,600 \times 4\,400$ & 3.80 & 3.0 & \\ 
LBS18-MM2   & 05:46:28.5 &  00:21:41 	& 550 &$12\,000 \times 8\,400$ & 1.40 & 4.0 & \\ 
LBS18-MM3   & 05:46:29.5 &  00:20:16 	& 790 & $7\,200 \times 7\,200$ & 2.00 & 3.0 & \\ 
LBS18-MM4   & 05:46:30.0 &  00:19:52 	& 250 & $4\,800 \times 3\,200$ & 0.65 & 3.0 & \\
LBS15-MM4   & 05:46:37.7 &  00:27:04 	& 150 & $6\,400 \times 1\,600$ & 0.40 & 3.0 & \\ 
\hline
LBS8-MM1    & 05:46:55.0 &  00:23:25 	& 500 &$12\,000\times 10\,400$ & 1.25 & 3.5 & \\ 
LBS8-MM2    & 05:46:57.1 &  00:20:10 	& 490 & $8\,400 \times 5\,200$ & 1.25 & 3.0 & \\
LBS8-MM3    & 05:46:57.1 &  00:23:56 	& 320 & $4\,800 \times 4\,400$ & 0.80 & 3.5 & \\ 
LBS8-MM4   & 05:46:58.2 &  00:20:12 	& 210 & $4\,800 \times 1\,600$ & 0.55 & 3.0 & \\ 
LBS8-MM5   & 05:46:58.4 &  00:24:34 	& 150 & $4\,800 \times 2\,000$ & 0.40 & 3.5 & \\ 
LBS8-MM6   & 05:46:59.8 &  00:20:26 	& 160 & $5\,200 \times 4\,000$ & $<0.40$ & 3.0 & line contamination of $S_{850}^\mathrm{~int}$\\ 
LBS8-MM7   & 05:47:00.3 &  00:20:35 	& 180 & $4\,000 \times 3\,200$ & $<0.45$ & 3.0 & line contamination of $S_{850}^\mathrm{~int}$\\ 
LBS8-MM8    & 05:47:01.0 &  00:26:19 &$2\,100$&$12\,000 \times 6\,800$ & 5.30 & 3.0 & \\ 
LBS8-MM9   & 05:47:01.1 &  00:20:37 	& 130 & unresolved	       & $<0.35$ & 3.0 & line contamination of $S_{850}^\mathrm{~int}$\\ 
LBS8-MM10    & 05:47:01.7 &  00:18:03 &$1\,140$&$12\,800 \times 7\,600$ & 2.85 & 3.5 & \\ 
LBS8-MM11    & 05:47:02.0 &  00:20:45 	& 250 & $4\,000 \times 1\,600$ & $<0.65$ & 3.0 & line contamination of $S_{850}^\mathrm{~int}$\\ 
LBS8-MM12   & 05:47:02.7 &  00:22:55 	& 270 & $5\,200 \times 1\,600$ & 0.70 & 3.5 & \\ 
LBS8-MM13  & 05:47:03.2 &  00:19:40 	& 110 & unresolved	       & 0.30 & 4.0 & \\
LBS8-MM14   & 05:47:03.3 &  00:22:35 	& 400 & $3\,600 \times 1\,600$ & 1.00 & 3.5 & \\
LBS8-MM15  & 05:47:03.6 &  00:19:54 	& 110 & unresolved	       & 0.30 & 4.0 & \\  
LBS8-MM16   & 05:47:03.6 &  00:20:08 	& 210 & $4\,400 \times 3\,200$ & 0.55 & 4.0 & \\ 
LBS8-MM17  & 05:47:04.2 &  00:20:23 	& 140 & unresolved	       & 0.35 & 4.0 & \\
LBS8-MM18   & 05:47:04.7 &  00:21:45 &$29\,000$&$9\,600 \times 8\,400$ & 9.20 & 3.5 & NGC2071-IRS envelope, $T_\mathrm{d}=40$~K\\ 
LBS8-MM19   & 05:47:06.4 &  00:22:36 	& 540 & $5\,600 \times 3\,600$ & $<1.35$ & 3.0 & line contamination of $S_{850}^\mathrm{~int}$\\ 
LBS8-MM20  & 05:47:07.5 &  00:22:46 	& 210 & $2\,800 \times 1\,600$ & $<0.55$ & 2.5 & line contamination of $S_{850}^\mathrm{~int}$\\ 
LBS8-MM21   & 05:47:07.9 &  00:25:07 	& 210 & $4\,800 \times 3\,600$ & 0.55 & -- & \\ 
LBS8-MM22  & 05:47:08.2 &  00:22:52 	& 220 & unresolved	       & $<0.55$ & 2.5 & line contamination of $S_{850}^\mathrm{~int}$\\
LBS8-MM23   & 05:47:10.4 &  00:21:15 &$1\,360$& $5\,600 \times 4\,000$ & 1.10 & 3.5 & SSV37 envelope, $T_\mathrm{d}=20$~K\\
LBS8-MM24  & 05:47:10.7 &  00:22:29 	& 110 & unresolved	       & 0.30 & 3.5 & \\
LBS8-MM25  & 05:47:11.9 &  00:22:45 	& 150 & unresolved	       & 0.40 & 3.5 & \\
LBS8-MM26  & 05:47:12.4 &  00:23:27 	& 150 & $4\,400 \times 1\,600$ & 0.40 & 3.0 & \\ 
LBS8-MM27   & 05:47:12.6 &  00:22:23 	& 530 & $7\,600 \times 4\,400$ & 1.35 & 3.0 & \\ 
\hline
LBS6-MM1   & 05:47:14.3 &  00:21:30 	& 125 & unresolved	       & 0.30 & 3.0 & \\ 
LBS6-MM2    & 05:47:16.0 &  00:21:24 &$1\,050$&$12\,400 \times 5\,600$ & 2.65 & -- & \\
LBS6-MM3   & 05:47:17.2 &  00:21:27 	& 150 & unresolved	       & 0.40 & -- & \\  
LBS6-MM4    & 05:47:24.9 &  00:20:59 &$1\,670$& $5\,200 \times 3\,600$ & 4.20 & 3.0 & \\ 
LBS6-MM5    & 05:47:25.1 &  00:18:49 	& 410 &$10\,400 \times 4\,800$ & 1.05 & 3.0 & \\ 
LBS6-MM6    & 05:47:26.2 &  00:19:56 	& 540 &$12\,800 \times 3\,600$ & 1.35 & 3.0 & \\ 
LBS6-MM7    & 05:47:26.5 &  00:20:45 	& 320 & $6\,400 \times 2\,400$ & 0.80 & 3.0 & \\ 
LBS6-MM8    & 05:47:29.8 &  00:20:39 	& 330 & $8\,400 \times 4\,400$ & 0.85 & 3.5 & \\ 
LBS5-MM1    & 05:47:32.5 &  00:20:24 	& 610 & $7\,200 \times 3\,600$ & 1.55 & 3.5 & \\ 
LBS5-MM2    & 05:47:35.1 &  00:20:21 	& 260 & $8\,000 \times 3\,600$ & 0.65 & 3.5 & \\
LBS5-MM3    & 05:47:36.9 &  00:20:07 	& 760 & $5\,600 \times 4\,800$ & 0.60 & 3.5 & 05450+0019 envelope, $T_\mathrm{d}=20$~K\\
\hline 
LBS16-MM1   & 05:46:36.3 &  00:05:49 	& 500 & $7\,200 \times 6\,000$ & 1.25 & 3.5 & \\
LBS11-MM1   & 05:46:45.3 &  00:07:31 	& 540 & $6\,800 \times 4\,400$ & 1.35 & 3.5 & \\ 
LBS11-MM2   & 05:46:46.1 &  00:07:09 	& 580 & $8\,400 \times 5\,200$ & 1.45 & 3.5 & \\ 
LBS11-MM3   & 05:46:47.3 &  00:07:27 	& 420 & $6\,400 \times 4\,400$ & 1.05 & 3.5 & \\
LBS7-MM1    & 05:47:04.9 &  00:14:59 &$1\,020$&$16\,800 \times 9\,600$ & 2.55 & 3.5 & \\ 
LBS7-MM2    & 05:47:05.2 &  00:13:21 	& 200 &$10\,800 \times 1\,600$ & 0.50 & -- & \\ 
LBS7-MM3    & 05:47:06.7 &  00:12:35 &$1\,200$& $9\,600 \times 6\,400$ & 3.00 & 3.0 & \\ 
LBS7-MM4    & 05:47:15.3 &  00:18:42 	& 300 & $6\,000 \times 4\,400$ & 0.75 & 3.5 & \\
\hline 
\end{tabular}
\end{table*}

\addtocounter{table}{-1}
\begin{table*}[htbp]
\caption{continued. Characteristics of the submm condensations identified in NGC2068/2071}
\begin{tabular}{|lcc|rcc|rl|}
\hline
 Condensation & \multicolumn {2} {c} {Coordinates}~$^{(1)}$ &
$S_{850}^\mathrm{~int}$~$^{(2)}$ & {\it FWHM}~$^{(3)}$ & $M$~$^{(4)}$
& $\alpha_{450}^{850}$ & Comments\\
 Name & $\alpha_{2000}$ & $\delta_{2000}$ & [mJy] & [AU$~\times~$AU] &
[$M_\odot$] & (5) & \\
\hline\hline
LBS17-MM1   & 05:46:24.3 & -00:00:06 	& 460 &$13\,200 \times 5\,600$ & 1.15 & 4.0 & \\ 
LBS17-MM2  & 05:46:24.5 & -00:00:22 	& 160 & unresolved	       & 0.40 & 4.0 & \\ 
LBS17-MM3  & 05:46:24.6 & ~00:00:13 	& 300 & $7\,600 \times 4\,800$ & 0.75 & 4.0 & \\ 
LBS17-MM4   & 05:46:26.8 & ~00:01:07 &$1\,300$&$14\,800 \times 7\,200$ & 3.25 & 3.5 & \\ 
LBS17-MM5   & 05:46:27.6 & ~00:01:31 	& 300 & $7\,600 \times 7\,600$ & 0.75 & 3.5 & \\ 
LBS17-MM6   & 05:46:27.9 & -00:00:52 &$1\,710$& $6\,000 \times 4\,400$ & 4.30 & 3.5 & \\ 
LBS17-MM7   & 05:46:28.1 & -00:01:39 	& 760 & $8\,800 \times 7\,200$ & 1.90 & 3.5 & \\ 
LBS17-MM8   & 05:46:30.1 & -00:01:07 	& 340 & $6\,800 \times 2\,400$ & 0.85 & 3.5 & \\ 
LBS17-MM9   & 05:46:31.0 & -00:02:35 &$2\,400$& $5\,600 \times 5\,600$ & 3.00 & 3.0 & LBS17-H envelope, $T_\mathrm{d}=15$~K\\ 
LBS17-MM10   & 05:46:32.3 & -00:00:40 	& 330 & $8\,800 \times 1\,600$ & 0.85 & 4.0 & \\ 
LBS17-MM11  & 05:46:32.9 & -00:00:23 	& 125 & unresolved	       & 0.30 & 4.0 & \\ 
LBS17-MM12 & 05:46:33.4 & -00:00:12 	& 180 & unresolved	       & 0.45 & 4.0 & \\
LBS17-MM13 & 05:46:33.5 & -00:00:02 	& 160 & unresolved	       & 0.40 & 4.0 & \\

LBS10-MM1   & 05:46:34.9 & ~00:00:32 	& 250 & unresolved	       & 0.65 & 4.0 & \\
LBS10-MM2   & 05:46:35.8 & ~00:00:33 	& 260 & $5\,200 \times 4\,000$ & 0.65 & 3.5 & \\
LBS10-MM3   & 05:46:37.7 & ~00:00:34 	& 990 & $8\,800 \times 4\,400$ & 2.50 & 3.5 & \\
LBS10-MM4  & 05:46:39.1 & ~00:00:33 	& 120 & unresolved	       & 0.30 & 4.0 & \\
LBS10-MM5   & 05:46:39.4 & ~00:01:11 	& 940 & $5\,600 \times 4\,400$ & 2.35 & 3.5 & \\
LBS10-MM6   & 05:46:40.5 & ~00:00:34 	& 360 & $9\,200 \times 5\,600$ & 0.90 & 4.0 & \\
LBS10-MM7   & 05:46:43.0 & ~00:00:47 &$1\,400$& $9\,200 \times 9\,200$ & 3.50 & 4.0 & \\
LBS10-MM8   & 05:46:45.1 & ~00:00:18 	& 230 & $6\,400 \times 5\,200$ & 0.60 & 4.0 & \\
LBS10-MM9   & 05:46:47.4 & ~00:00:25 &$1\,710$& $6\,400 \times 4\,800$ & 1.35 & 3.5 & 05442-0000 envelope, $T_\mathrm{d}=20$~K\\
LBS10-MM10   & 05:46:48.4 & ~00:01:31 	& 410 & $5\,200 \times 5\,200$ & 1.05 & 4.0 & \\
LBS10-MM11  & 05:46:48.9 & ~00:01:22 	& 370 & $9\,600 \times 4\,800$ & 0.95 & 4.0 & \\
LBS10-MM12 & 05:46:49.7 & ~00:00:20 	& 165 & unresolved	       & 0.40 & 4.0 & \\ 
LBS10-MM13  & 05:46:49.7 & ~00:02:04 	& 540 & $9\,200 \times 3\,600$ & 1.35 & 4.0 & \\ 
LBS10-MM14  & 05:46:50.7 & ~00:02:07 	& 440 & $8\,800 \times 4\,000$ & 1.10 & 3.5 & \\
LBS10-MM15 & 05:46:51.4 & ~00:00:00 	& 150 & unresolved	       & 0.40 & 4.0 & \\ 
LBS10-MM16  & 05:46:52.8 & ~00:01:48 	& 150 & unresolved	       & 0.40 & 4.0 & \\ 
LBS10-MM17  & 05:46:54.2 & -00:00:24 	& 310 &$10\,400 \times 3\,600$ & 0.80 & 3.5 & \\ 
LBS10S-MM1  & 05:46:43.1 & -00:01:42 	& 230 &$17\,200 \times 4\,000$ & 0.60 & 4.5 & \\ 
\hline
\end{tabular}
\begin{list}{}{}
\item[ (1) ] {The absolute positional accuracy is better than 
$\sim 5\arcsec$.}
\item[ (2) ] {Integrated flux at $850\,\mu$m estimated (after 
background subtraction) over an area twice the size of col.~[5] when
convolved with the 13$\arcsec$ beam.  The absolute calibration
uncertainty is $\sim 20\%$.  The typical rms noise is $\sim 22~{\rm
mJy} \times \sqrt{({\rm \it FWHM}^2+{\rm \it HPBW}^2) /{\rm \it
HPBW}^2} \times 2$.}
\item[ (3) ] {Deconvolved {\it FWHM} size derived from a 2D-Gaussian fit 
to the $850\,\mu$m map after background subtraction.}
\item[ (4) ] {Mass derived from the $850\,\mu$m integrated flux of
col.~[4].  Assumed dust temperature and opacity are
$T_\mathrm{d}=15$~K and $\kappa_{850} = \rm 0.01~cm^{2}\, g^{-1}$ for
starless condensations, $T_\mathrm{d}= 20-40$~K and $\kappa_{850} =
\rm 0.02~cm^{2}\, g^{-1}$ for protostellar envelopes (see also
Sect.~2).}
\item[ (5) ] {Spectral index $\alpha_{450}^{850}$ (where $S_\nu \propto
\nu^\alpha$) measured between $450\:\mu$m and $850\:\mu$m in 
an $18\arcsec$ beam at the position of each condensation {\it without
background subtraction}. The typical error bar on $\alpha_{450}^{850}$
is $\pm 0.5$. Since the background emission often dominates and has a
spectral index of $\sim $~4, we estimate that the intrinsic spectral
indices of the condensations are lower than the values listed in
col.~[7] by $\Delta \alpha \sim 1 $ on average.}
\end{list}
\end{table*}

\end{document}